\documentstyle[multicol,aps,prl,psfig]{revtex}
\input{epsf.tex}
\begin{document}

\title{Attractive Interactions Between Rod-like Polyelectrolytes:
Polarization, Crystallization, and Packing}

\author{ Francisco J.~Solis and Monica Olvera de la Cruz}

\address{ Department of Materials Science and Engineering, \\
	 Northwestern University, Evanston, Illinois 60208-3108. }

\maketitle
\begin{abstract}
	We study the attractive interactions between rod-like 
charged polymers in solution that appear in the presence of
multi-valence counterions. 
The counterions condensed to the rods exhibit both a strong 
transversal polarization and a longitudinal crystalline arrangement.  
At short distances between the rods, the fraction of condensed
counterions increases, and the majority of these 
occupy the region between the rods, where they minimize their 
repulsive interactions by arranging themselves into  packing 
structures. The attractive interaction is strongest for multivalent 
counterions. Our model takes into account
the hard-core volume of the condensed counterions and their angular 
distribution around the rods. The hard core constraint strongly 
suppresses longitudinal charge fluctuations.

{PACS numbers: 61.20.Qg, 61.25.Hq, 87.15.Aa }
\end{abstract}

\begin{multicols}{2}

%%%%%%%%%%% 1 %%%%%%%%%%%

	Strongly charged polymers precipitate from a dilute solution
into compact structures when high-valence counterions (oppositely charged 
particles) are added to the solution \cite{Widom,DNA1,DNA2,Syn,virus,Raspaud}.
The counterions experience strong electrostatic attractions 
to the backbone of the chains, and a finite fraction of them
``condense'', i.e., are found within short distance from the 
chains \cite{Manning}. 
Counterions are more attracted to compact chains or 
aggregates of rod-like chains.
This creates the possibility of a transition from 
single chains with a small number of condensed counterions to 
almost neutral aggregates of chains, or even mono-molecular collapse
in the case of flexible polymers. 
These aggregates are stable only when the internal arrangement of the
counterions within them, provides a strong enough cohesive
energy. 

%%%%%%% 2 %%%%%%%%%%%%%%%

In this letter we study the attraction between two rod-like
polyelectrolytes. We show that it is essential to include the 
size and angular degrees of freedom (around the rods) of the counterions
as well as the discrete nature of the charge along the polyelectrolytes to 
find the origin and strength of the counterion mediated attraction.
Our work suggests that these factors are also crucial in determining 
the collapse of flexible and semi-flexible polyelectrolytes 
 recently studied in Refs.[8-12].
Experimental observations show that the size of the precipitating particles 
is indeed a relevant parameter in the problem \cite{Widom,Raspaud}.

%%%%%%% 3a %%%%%%%%%%%%%%%

	It has been argued that longitudinal charge fluctuations
resulting from the thermal motion of point counterions induce
attractions between rod-like polyelelctrolytes \cite{Barrat,Ha} and induces
"buckling" of semi-flexible polyelelctrolytes \cite{kardar,Hansen}.  
Here, we show that such charge fluctuation are
suppressed when the hard core volume of monomers and counterions are
taken into account. Instead, we find that the counterions arrangement 
around the rods create a non-zero transversal polarization as the distance 
between the chains decreases. At very short distances between the rods we 
find strong longitudinal correlations but only at very short
wave-lengths, implying a crystalline state along the rod, reinforcing, for
the case of multivalent counterions, the attractive interactions due to
polarization.  

%%%%%%%%%%%% 3b  %%%%%%%%%%

The crystalline structure of the counterions when 
the rods are at short distance from each other has been 
observed in simulations by Gr{\o}nbech-Jensen {\it et al.}
 \cite{Jensen}, and has been theoretically proposed by 
Arenzon {\it et al.} \cite{Levin} and Shklovskii \cite{Sh}. 
These previous  theoretical works retain some of the 
small size effects of a realistic system, but again use the  assumption of 
negligible size counterions. These models do not reveal the polarization
effects that appear when the angular degrees of freedom around the rod are 
considered. 

%%%%%%%%%%%%%%%%%%%% fig 1 %%%%%%%%%%%%%%%%%%%%%%%%%%%%

\begin{center}
\begin{minipage}[H]{3.2in}
\epsfxsize=3.0in \epsfbox{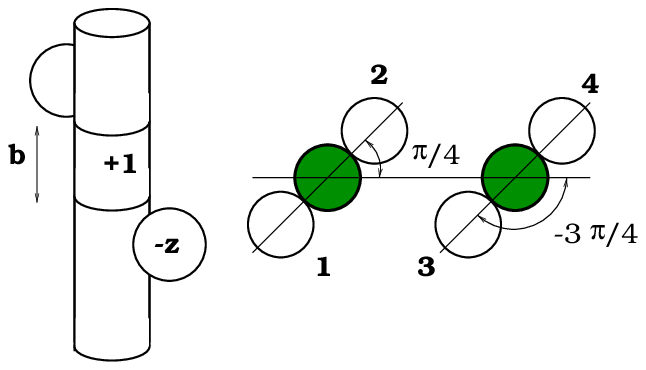}
\begin{figure}
\caption{The diameter of the rod, the size of the counterions and 
the basic spacing between charges all have  magnitude $b$. 
We use two representative states
for each monomer in each rod. All occupation is assumed to take
place parallel to the centers of the monomers, so that the possible 
places for occupation form the lattices $1,2,3$, and $4$, here shown 
from the top.}
\end{figure}
\end{minipage}
\end{center}

%%%%4%%%%

The polymer chains are modeled as rigid rods formed by $N$ repeat units,
each with charge $+1$ centered at each of the monomers,
as shown in Fig.~1.
We label the pair of rods as $A$ and $B$.
The diameter of the rods and the spacing between charges are both $b$.
The total length of the chains is $L=Nb$. We will only consider the case in 
which the chains are parallel 
and separated by a distance $r$ measured from center to center. The 
counterions carry a charge $-Z$, and for simplicity  we take them 
to be of diameter $b$ as well. Overall charge neutrality of the system 
implies that there are $N/Z$ counterions per chain. We assume a 
concentration of chains $c$,
to be such that the average distance between rods (if non-interacting) is also 
$L$, so that $c=1/L^{3}$.

%%%%%5%%%%

A layer of condensed counterions surrounds each chain.
The location of a particular counterion is given by its position along the
rod $z$, a radial distance $r$ from the rod, and its angular
position $\theta$. We restrict the $z$ coordinate of the condensed counterions
to take values coinciding with those of the centers of the monomers. 
The radial distance form the rod for all condensed 
counterions is assumed to take always the  value $r=b$.  The angular 
variable is very important as it carries information 
about the local polarization of the rod-counterions system. A suitable 
simplification that retains this information consists on collapsing 
the range of angular positions to only two specific locations. These 
positions are located at angles of $\pi/4$ and $-3\pi/4$ with respect to 
the plane that contains both rods, and are labeled as shown in Fig.~1. 
These ad-hoc positions are meant to represent locations
of counterions on one rod that look towards or away from the other 
rod, and have the convenience of allowing  us to 
consider distances between the centers of the rods as small as $\sqrt{2}b$
without worrying about overlapping of the counterions that  would require     
explicit introduction of hard-core repulsions. The choice of 
the angles does not significantly change the results. 
In short, we have modeled the condensed layers
as  four linear lattices parallel to the rods, whose sites can be occupied
by the counterions.

%%%%%6%%%%

Since our selected geometry takes care of the hard-core interactions 
between the particles of the system, we can construct a Hamiltonian with only 
electrostatic interactions. This is given by:
%%%%%%%%%%%%%%%%%%  (1)  %%%%%%%%%%%
\begin{equation}
	H_{c}=	{\it l}_{B}
	\sum_{s\neq t}\frac{1}{2}Z^2
	\frac{1}{|{\bf r}_{s}-{\bf r}_{t}|}
	-{\it l}_{B}\sum_{s}Z\phi({\bf r}_{s}).
\end{equation}
%%%%%%%%%%%%%%%%%%%%%%%%%
 We measure energies in units of $k_{B}T$, and the prefactor $\it{l}_{B}$ is
the dimensionless ratio of the Bjerrum length to the monomer size 
$e^2/\varepsilon b k_{B}T$,
with $e$ being the electron charge, $k_{B}$ the Boltzmann constant $T$ the
temperature and $\varepsilon$ the dielectric constant of water. $\phi$ is the 
electrostatic potential created by the charged rods. Ignoring end effects 
from the rods, this  potential is well approximated by 
$\phi({\bf r})=2{\it l}_B (\ln(L/r_{A})+\ln(L/r_{B}))$, where $r_{A}, r_{B}$,
are the distances from the point ${\bf r}$ to the axis of the rods $A$ 
and $B$, respectively. 
For the condensed counterions, it is better to change to a local
charge representation.
Each site of the four lattices
can be occupied by one counterion (we neglect multiple occupation), 
and thus it will carry a charge $q_{i}(n)$ that can be either $-Z$ or zero. 
The index $i$ is the is the (longitudinal) position in the lattice, that can 
range from $1$ to $N$. 
%%%%%% 7 %%%%%%%%%%%

	Since the number of monomers is large, we expect that the 
total number of condensed counterions for a given inter-rod separation
will have a narrowly peaked probability distribution. We 
construct a free energy that assumes that the number of counterions 
condensed to each of the lattices is fixed, and then we will find the minimum 
with respect to the occupation numbers. Lattice $i$ carries a fraction $f_{i}$
of the number of counterions per rod $N/Z$, and the fraction condensed to rod
A is  $f_{A}=f_{1}+f_{2}$, etc. 

%%%%%%%% 8 %%%%%%%%%%

	The free counterions form a 
dilute charged gas that occupies a volume $V=2L(L^2-4b^2)$. 
Because of their low density their contribution to the free energy from 
correlations and screening is negligible. A  free counterion at a distance 
$r_{c}$ from the center of the system feels a potential  
given approximately by 
$\phi=2 ({\it l}_{B}/e)(2-f_{A}-f_{B}) (1-(r_{c}^{2}/2L^{2}))\ln (L/r_{c}) $, 
that arises from the effective (uncompensated) charge 
of the two rods and a cylindrical shell of uniformly distributed 
free counter-ions. Averaging this potential over the volume, and 
adding the entropic contribution, we obtain the free energy per monomer 
due to the free counterions:
%%%%%%%%%%%% (2) %%%%%%%%%%%%%%
\begin{eqnarray}
	F_{f}&=& \frac{1}{2Z}(2-f_{A}-f_{B})
	\left(\ln( (2-f_{A}-f_{B})Nb^{3}/V)-1\right) \nonumber \\
	&&+\frac{1}{2}(2-f_{A}-f_{B})^{2}
		\left(-\frac{3}{2}{\it l}_{B} \right) 
\end{eqnarray}
%%%%%%%%%%%%%%%%%%%%%%%%%%%%%%%%

%%%%%%%%%%%%%%%%%%% 9 %%%%%%%%%%%%%%%%%%%%

	To obtain the contribution from the condensed counterions,
we consider first a high-temperature approach in which we add fluctuations
to a uniformly distributed state. 
Given a condensed fraction $f_{i}$ at the lattice $i$, there are
$f_{i}N/Z$ occupied sites. 
The statistical sum over all states satisfying this restriction 
can be replaced at high temperatures by Gaussian integrations over 
a set of continuous local charge variables. The 
charge  $q_{i}(n)$ is represented by a density $\rho_{i}(n)$ with mean 
$-f_{i}$ and variance  $\sigma_{i}^{2}=Z^{2}(f_{i}/Z)(1-f_{i}/Z)$.
This form of the variance is consistent with our assumption of 
a maximum of one counterion per lattice site, as required by
our geometry and hard-core constraints \cite{correction}.
We can pass to a discrete Fourier representation of the local charge
of the form 
%%%%%%% (3) %%%%%%%%%%%%
\begin{equation}
	\rho_{i}(n)=-f_{i}+\sum_{k \neq 0}\rho_{i}(k)\exp(ikn)
\end{equation}
%%%%%%%%%%%%%%%%%%%%%%%
where the only Fourier modes considered are of the form $k= \pm 2\pi m/N$,
with $m$ ranging from $-N/2$ to $N/2$.
This transformation diagonalizes the part of the
Hamiltonian, Eq.~(1), that corresponds to the condensed counterions,
and leads to the free energy per monomer:  
%%%%%%%%%% (4) %%%%%%%%%%%%
\begin{eqnarray}
	F_{c}&=&\frac{1}{4}\sum_{i,j}f_{i}V^{0}_{ij}f_{j}
	+\frac{1}{4N}\sum_{k\neq 0}
		\ln \det\left[{\bf I}+{\bf S}{\bf V}(k) \right] \nonumber \\
	&&-\frac{1}{2}\sum_{i} f_{i}\phi_i +\frac{1}{2}\sum_{i}
		\frac{f_{i}}{Z}\left(\ln\frac{f_{i}Nb^{3}}{V_{c}}-1\right).
\end{eqnarray}
%%%%%%%%%%%%%%%%%%%%%%%%%%%%
The first term is the contribution from the zero modes where 
the interaction matrix between lattices is 
$V^{0}_{ij}=2{\it l_{B}}\ln(L/r_{ij})$ with $r_{ij}$ the distance between 
the axis of the lattices, and the diagonal terms are given by 
$V^{0}_{ii}=2{\it l_{B}}\ln(L/b)$. 
The second term is the result of the Gaussian integration over fluctuations, 
where ${\bf I}$ is the identity matrix, ${\bf S}$ is a diagonal matrix with 
entries $\sigma_{i}^{2}$, 
and the interaction terms for the Fourier modes are of the form 
$V_{ij}(k)=2 {\it l}_{B}K_{0}(kr_{ij})$, 
and $V_{ii}(k)=-2 {\it l}_{B}ci(kb)$,
with $K_{0}$ the modified Bessel function, and $ci$ the cosine integral 
function. The integrals that define the diagonal expressions are evaluated 
using the lattice spacing $b$ as short distance cutoff so that there is 
no need for self-energy 
subtractions. The finite sum over modes also corresponds to 
consider distances larger than one lattice spacing. The last term is the 
entropic contribution of placing ${f_{i}N/Z}$ counterions in a volume 
$V_{c}= Lb^{2}$. 

%%%%%%%% 10 %%%%%%%%%%%%%%%

	{\it A posteriori} we have found that at short distances between 
the rods the condensed counterions are almost fully polarized, occupying the 
states that face towards the other rod. The fraction of condensed charge
approaches $1$ in these sites, while the outward states are almost completely
depleted. We can look then in more detail at the calculation of 
the free energy for the case of almost full occupation of one of the lattices 
$f\rightarrow 1$, so that  
the variance of the fluctuations becomes $\sigma^{2}\approx (Z-1)$.
The diagonal interaction term for the highest Fourier modes 
$k\approx \pm \pi/b$ is negative and
of order ${\it l}_{B}$, corresponding to a decrease in energy from 
vacating a state next to an occupied one. Since typical values 
for ${\it l}_{B}$
at room temperature are always larger than $1$, these diagonal 
terms dominate the interaction matrix ${\bf V}$, and therefore the matrix 
${\bf I}+{\bf SV}$ has diagonal elements $1-(Z-1){\it l}_{B}$. 
When the condition 
%%%%%%%%%%%%%%%%% (5) %%%%%%%%%%
\begin{equation}
	(Z-1){\it l}_{B} > 1
\end{equation}
%%%%%%%%%%%%%%%%%%%%%%%%%%%%%%%%%%
is satisfied, the matrix acquires negative eigenvalues, making the 
determinant divergent and the high-temperature approximation incorrect. 
Clearly, this occurs in most cases, except for mono-valent counterions
($Z=1$) or for very weakly charged polyelectrolytes for which 
${\it l}_{B} \ll 1$. 
The divergence in the 
determinant for multi-valent counterions signals the onset of 
crystallization, and thus the  free energy should be calculated on
the basis of a dominant crystalline ground state.  
Once we use the correct ground state for both mono- and multi-valence cases, 
it can be shown that the corrections from fluctuations are very small,
that is, the contribution of the determinant in Eq.~4 
becomes negligible once the divergent modes are subtracted.

%%%%%%%%%%% 12 %%%%%%%%%%%%%%%%%

The proper ground state for each of the inner lattices, when approaching 
full condensation, is clearly given by an arrangement 
in which counterions are placed one in every $Z$ sites.
(This is the case already for $Z=1$.)
As the rods approach each other, the inner lattices interact 
strongly, but they do not destroy the ground state arrangement. 
Instead, they can choose a location of the occupied sites, so as to minimize
their repulsive interaction. For $Z=2$, for example, one expects
one of the internal lattices to be filled in the even sites, while the second
in the odd ones.

%%%%%%%%%%%  13 %%%%%%%%%%%%%%%%%

The calculation of the free energy for the multi-valence case can 
be carried out using the results for the mono-valent case which are 
given by Eq.~(4). We 
simply renormalize the lattice spacing to $Zb$ and reduce the available 
sites by a factor of $Z$. The elements of the interaction matrix for 
the ground states 
are now given by $V^{m}_{ij}=2 {\it l}_{B}\ln (L/r'_{ij})$, with a modified 
distance between the lattices $r'_{ij}=(r_{ij}^{2}+(bZ/2)^2)^{1/2}$ 
that takes into account the mismatch between the occupied  sites  
in the lattices. The diagonal term is simply 
$V_{ii}^{m}=2 {\it l}_{B}\ln (L/Zb)$. 

%%%%%%%%%%%% 14 %%%%%%%%%%%%%%

In Fig.~2 we present the calculated free energy of the system as a 
function 
of the distance between rods. The numerical values for the constants
of the system are $T=300$, $b=1.8A^{o}$, so that ${\it l}_{B}=4.1$,
and $N=10^5$. There is a well of  attraction for 
the mono-valent case of about $.5 kT$, which is not enough to bind the 
rods, and further, the local minimum at short distances turns out
to be of higher energy than the self-energy of two rods separated by
a distance $L$. For $Z=2$ and $3$, the depth of the well is of the order 
of $k_{B}T$, and the energy there is lower than their respective 
reference states.

%%%%%%% fig 1 %%%%%%

\begin{center}
\begin{minipage}[H]{3.2in}
\epsfxsize=3.0in \epsfbox{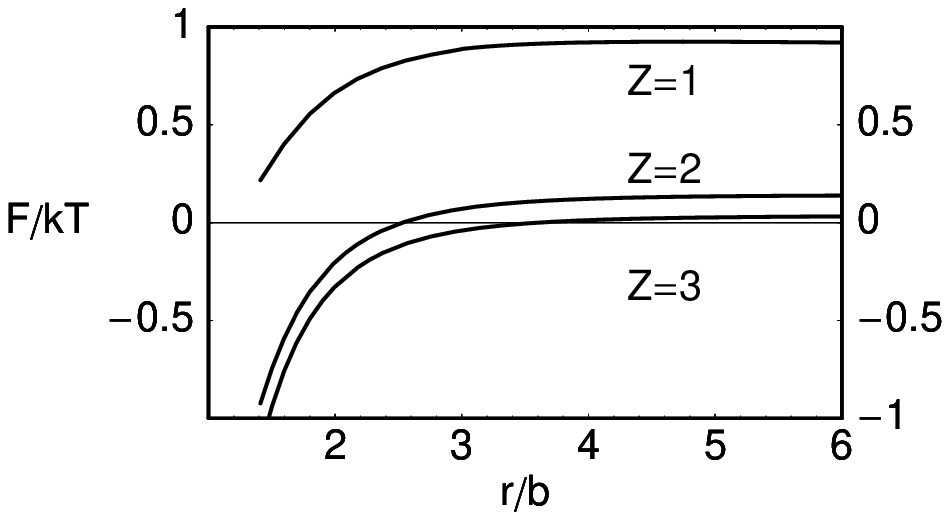}
\begin{figure}
\caption{ Results of the minimization of the free energy,
for parameters such that ${\it l}_{B}=4.1$, for valences $Z=1,2$, and $3$. 
Energies are shown in units of $k_{B}T$ and distances in units of $b$. 
The zero of the energy in each case is chosen to match the 
energy at large separations.}
\end{figure}
\end{minipage}
\end{center}

%%%%%%%%%%%%% 15 %%%%%%%%%%%%%%%

	Fig.~3 presents the total amount of condensed counterions for 
one of the  rods, $f_{A}=f_{1}+f_{2}$. This is always very close to $1$ 
for short distances and 
reaches a value near the Manning 
limit \cite{Manning},
$f= (1- 1/Z{\it l}_{B})$ at large separations. We measure the overall 
polarization $p$ of the rods by the ratio 
of the difference between the occupation of the inward and outward
positions to the total amount of condensed charges, thus for rod $A$:
%%%%%%% (6) %%%%%%%%%%%%%%%%
\begin{equation}
	p_{A}=\frac{f_{2}-f_{1}}{f_{2}+f_{1}}
\end{equation}
%%%%%%%%%%%%%%%%%%%%%%%%%%%%%
Results for the amount of polarization are shown in Fig.~4. 
It is clear that the charge will be perfectly balanced when the 
presence of the second rod is not felt. On the other hand, when the 
chains are in close contact is natural for the counterions to occupy the 
inner sites to be able to
interact with the positive charges of both rods, even at the expense 
of interacting with other condensed counterions. What it is surprising is 
that both the polarization and the extra condensation do not decay quickly, 
and it is necessary to 
set the distance between the rods to its maximum value $L$ to recover
the Manning limit and a symmetric state. 
A good test of the validity of this theory will be 
the measurement in simulations and experiments 
of the transversal polarization of the rods. 

\begin{center}
\begin{minipage}[H]{3.2in}
\epsfxsize=3.0in \epsfbox{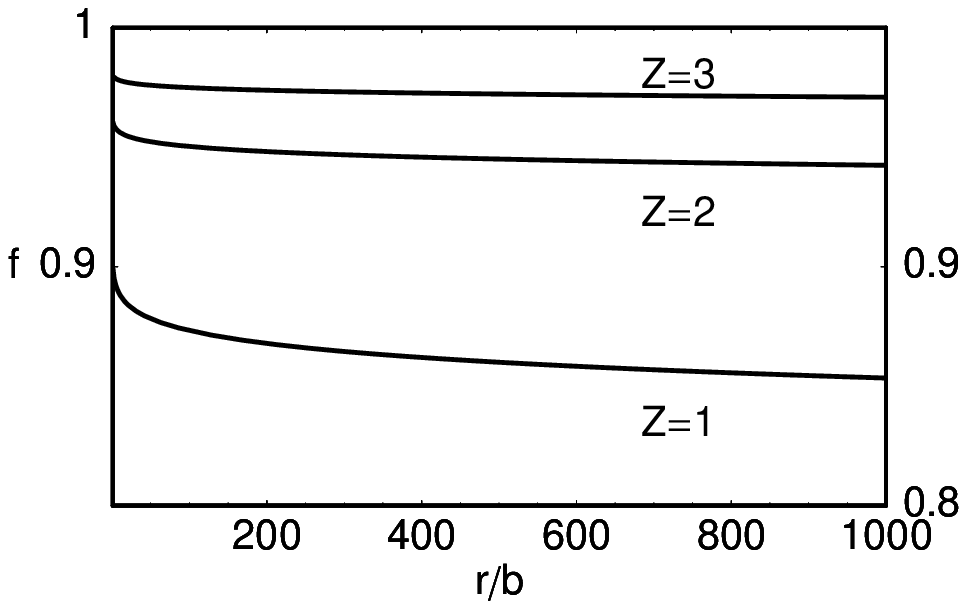}
\begin{figure}
\caption{The total fraction of condensed counterions to each of the rods, 
as a function 
of the separation between rods for valences $Z=1,2$, and $3$. The scale in 
which 
the  condensed fraction decays to a single rod value is of the order of
the size of the rod  $L=10^{5}b$. Note the change of scale 
with respect to Fig.~2. }
\end{figure}
\end{minipage}
\end{center}

\begin{center}
\begin{minipage}[H]{3.2in}
\epsfxsize=3.0in \epsfbox{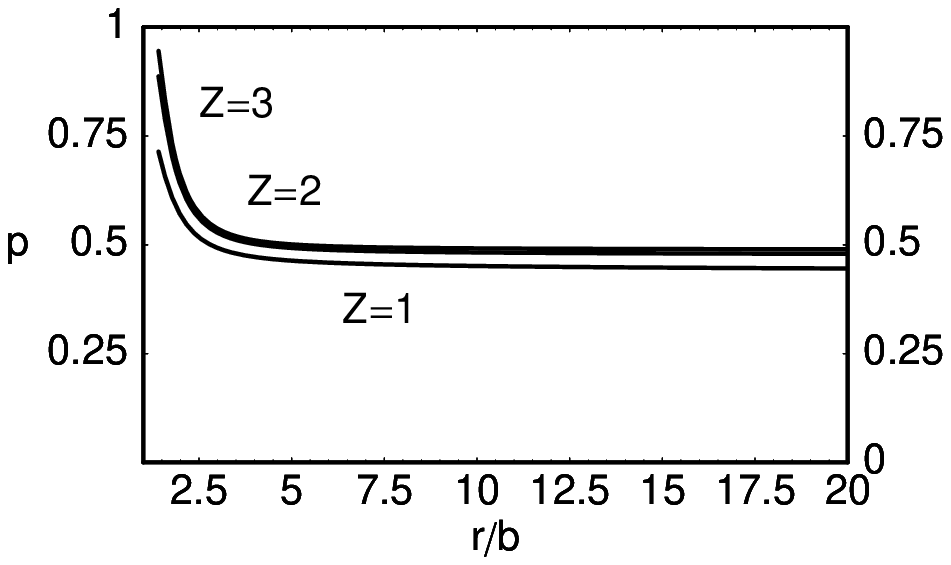}
\begin{figure}
\caption{ Polarization of the rods as a function of the separation 
between rods for valences $Z=1,2$, and $3$. At very short distances, the 
polarization is almost complete. Away from the region of strong
attraction, the polarization still remains important and will
decay to zero only at distances of order $L$. The plots for $Z=2,3$,
overlap almost completely.}
\end{figure}
\end{minipage}
\end{center}

%%%%%%%%%%%%%% 17 %%%%%%%%%%%%%%

	In summary, we have shown that the interaction between two charged 
rod-like polymers generates a strong transversal polarization of their 
condensed charges and that at short distances the two rods are strongly driven 
towards higher counterion condensation. This forces the counterions to 
crystallize and then to organize their respective crystals into a packing 
structure. The final result is an important attraction between 
the rods when the counterions are multi-valent. We found that the finite size 
of the counterions and their angular degrees of freedom are essential to 
determine the nature and strenght of the counterion mediated attractions
in rigid-rods, and we expect this to be also the case in flexible and
semiflexible polyelectrolytes.

\section*{ACKNOWLEDGMENTS}

	This work was sponsored by the National Science Foundation,
grant DMR9807601.

%%%%%%%%%%

%%%%%%%%%%%
\end{multicols}


\begin{thebibliography}
\bibitem{correction} Our variance differs from the expression
used in models that assume point-size counterions, refs.\cite{Ha,kardar},
where the variance is proportional to $f_{i}$. A variance proportional
to $f_{i}$ implies, at large condensation fractions, frequent multiple 
occupation at a monomer site that is inconsistent with the reduced 
space available to the counterions between the rods.
\end{thebibliography}
\end{document}